\documentclass[12pt,a4paper]{article}
\usepackage[german, english]{babel}
\usepackage{a4wide}
\usepackage{amsmath}
\usepackage{amssymb}
\usepackage{ifthen}
\usepackage{epsfig}

\newcommand{\rf}[1]{(\ref{#1})}

\newcommand{\beq}{\begin{equation}}
\newcommand{\eeq}{\end{equation}}
\newcommand{\beqr}{\begin{eqnarray}}
\newcommand{\eeqr}{\end{eqnarray}}

\newcommand{\lb}[1]{\label{#1}}

\newcommand{\bc}{\begin{center}}
\newcommand{\ec}{\end{center}}
\newcommand{\ct}[1]{\cite{#1}}
\newcommand{\bi}[1]{\bibitem{#1}}

\begin{document}
\begin{titlepage}
\title{Bistability in quantum nonlinear oscillator excited by stochastic force}

\author{I.E~Protsenko$^{1,2,3}$, E.D.~Protsenko$^{3}$ and A.V.~Uskov$^{1,2}$ \\ [4mm]
{\it {\footnotesize
$^1$P.N.~Lebedev Physical Institute, 119991, Russia, Moscow, Leninsky prospect 53
}}\\
{\it {\footnotesize
$^2$Advanced Energy Technologies Ltd., 143025, Russia, Moscow region, Skolkovo, Novaya str.~100.
}}\\
{\it {\footnotesize
$^3$National Nuclear Research University MEPhI, 115409, Russia, Moscow, Kashirskoe highway, 31.
}}\\
\date{~}
}
\maketitle
\begin{abstract}
We present approximate analytical method of analysis of stationary states of nonlinear quantum systems with the noise. As an example we consider quantum nonlinear oscillator excited by fluctuating force and found parameter regions with more than one stationary solutions. Existence of such region is the necessary condition for bistability. We neglect by fluctuations in the amplitude of oscillations but do not neglect by fluctuations in its phase. Then oscillator noise power spectrum depends on oscillator mean energy $n$, which leads to nonlinear integral equation for $n$. Analytical solution of this equation can be found.  Stationary states  of the oscillator are found for various spectrums of fluctuations of the exciting force. Linear stability analysis of stationary states was carried out.
\end{abstract}
\bigskip


\end{titlepage}

\section{Introduction}
Dynamics of nonlinear systems excited by fluctuating forces attracts attention for a long time. Well-known example of interesting phenomena in such systems is a stochastic resonance, when bistable system is excited by the stochastic and the regular forces together \ct{1,2,3}. There an increase in the signal to noise ratio, stochastic synchronization of switches between states of the system are possible; for the maximum amplification of the regular component of the signal optimum level of noise is found. Increase of the signal-to-noise ratio (SNR) of an amplifier is, obviously, very important practical problem.

Well-known nonlinear quantum systems, where regular (coherent) dynamics appears at noisy (incoherent) pump are lasers and related devices \ct{4,5}. In particular, the nonlinearity, i.e. saturation of lasing transition, leads to the narrowing of the laser linewidth with the increase of the intensity of the incoherent pump. General problem of locking of a self-oscillator (Van-der-Pol oscillator) by a random signal  related with lasing was discussed in \ct{05}. Optical laser systems provides convenient tools for studies of nonlinear dynamics with noise.
For example, enhancement in the output SNR and a noise-induced switching were predicted and experimentally observed in the three-level atomic optical bistability (AOB) systems \ct{06} -- \ct{010}.

Correct theoretical description of nonlinear systems excited by noise sources is a difficult task. For example,  one can't solve explicitly  Fokker-Planck equation and find correlation functions and spectral densities for nonlinear bistable classical oscillator \ct{1}. Many interesting questions concerning noise-induced transitions in nonlinear systems are discussed in \ct{5a}. For the case of lasers Schawlow-Townes formula \ct{4}  can't describe the linewidth of many kinds of them. Expressions different from that formula were derived for  "bad cavity" lasers (i.e. lasers with low quality cavities) \ct{6,7}, including plasmonic nanolasers \ct{8,9}; for "thresholdless" lasers (with high spontaneous emission at the threshold) \ct{10}. However general and relatively simple method of calculation of  linewidth of any laser, also with high degree of spontaneous emission noise, is not yet developed. Thus the development of simple and reliable method of treatment of quantum nonlinear systems with high degree of noise, also with the noise in the pump, is a topical problem.

In \ct{9,10} we presented approximate analytical method of calculation of stationary states and the linewidth of a laser below, at and above the lasing threshold, valid also for "bad cavity" lasers. The method is based at quantum Langevin equations and uses the assumption that fluctuations of populations of the lasing active medium are much smaller than fluctuations of laser medium dipole momentums and the field.
This is good assumption, in particular, for "bad cavity" lasers with high noise in the lasing mode and in the polarization of lasing medium, as it was confirmed in \ct{10} by numerical calculations. Usually quantum Langevin equations for fluctuations are obtained by linealization around the steady state \ct{10a}, so that the steady state does not depend on the noise. In our method the steady state does depend on the noise, in fact, the energy of the system is fully provided by the noise source.

The main purpose of this paper is to demonstrate  how the method of \ct{9,10} can be applied, at first approximation, to general quantum nonlinear oscillating systems, not necessary lasers. For that we use an example of nonlinear quantum oscillator excited by random force originated from the oscillator -- bath interaction. It is well-known that classical nonlinear oscillator  excited by regular force near the resonance has bistability in its stationary states \ct{11}. The necessary condition for the bistability is the existence of more than one stationary solution in some region of parameters. Here we restrict ourselves by finding of such parameter regions. By the method of \ct{9,10}  and following the approach of catastrophe theory \ct{12} we'll find areas of parameters with several (three) stationary states of quantum oscillator excited by random force with various spectra of fluctuations. Note that random force does not lead to a true bistability, even if parameter regions with many stationary solutions exist \ct{15}, \ct{16}: fluctuations of the oscillator energy and phase can lead to switching between stationary states stable without fluctuations. Thus the stationary states are metastable and there is only "quasi-bistability" at prescience of fluctuating force. We leave calculations of the oscillator energy fluctuations, detail analysis of stability conditions and switching dynamics (switching times, lifetimes of metastable states) in our system for the future.

In the first Section we derive equation for determining the mean energy of nonlinear quantum oscillator. In Section 2 we solve this equation in some particular cases and find regions of parameters with many stationary solutions. Results are summarized and discussed in Conclusion.

\section{Energy of oscillator excited by random force}

Classical equation of motion for coordinate $x$ of anharmonic oscillator is \ct{11}:
\beq
  \ddot{x} + 2\gamma\dot{x} + \omega_0^2x = F(t)\cos{(\omega_p t)}  - \beta x^3,   \lb{1_1}
\eeq
where $\gamma$ is a dumping rate, $\omega_0$ is a frequency of linear oscillations, $m$ is a mass, $\beta$ is the coefficient of nonlinearity. Nonlinear term $\alpha x^2$ presents in the right side of Eq.\rf{1_1} in \ct{11}; in order to simplify analysis we suppose  $\alpha = 0$.  In Eq.\rf{1_1}  $F(t) = f(t)/m$, $f(t)$ is an amplitude of external random force exciting (pumping) the oscillator.  Spectrum of fluctuations of $f(t)$ is centered at $\omega_p \approx \omega_0$. It may be that $\gamma  \ll \Gamma_p$ -- the half-width of the power spectrum of $f(t)$, however  $\Gamma_p \ll \omega_0$, i.e   $f(t)$ fluctuates slowly respectively to $\cos{(\omega_p t)}$.

Let us now consider quantum oscillator and suppose that $x(t)$ and $f(t)$  are operators, Eq.\rf{1_1} is Heisenberg-Langevin equation of motion for $x$. We replace $x$ in Eq.\rf{1_1} by bose-operator $ae^{-i\omega t}$
\beq
        x = \left(\frac{\hbar}{2m\omega_0}\right)^{1/2}(ae^{-i\omega_p t} + a^+e^{i\omega_p t}) \lb{1_2}
\eeq
and simplify Eq.\rf{1_1} using resonant approximation, assuming that  $a$ is changed slowly than  $e^{-i\omega_p t}$. In the left side of Eq.\rf{1_1} we take: $\omega_0^2- \omega_p^2 \approx -2\omega_0\delta$, where detuning $\delta = \omega_p - \omega_0 \ll \omega_0$, neglect $\ddot{a}$; neglect $\dot{a}$ in the term $\sim \gamma$ and leave only terms $\sim e^{-i\omega_p t}$. Thus we  obtain instead of Eq.\rf{1_1}:
\beq
    \dot{a} = i(\delta + ba^+a)a  - \gamma a + \sqrt{2\gamma}a^{in}(t),  \lb{1_3a}
\eeq
where normalized coefficient of nonlinearity $b = 3\hbar\beta/(8m\omega_0^2)$;  we carried out normal ordering of Boze operators in $x^3 \sim (ae^{-i\omega_p t} + a^+e^{i\omega_p t})^3$ and then re-defined oscillator frequency $\omega_0$.
In Eq.\rf{1_3a} the dumping term $-\gamma a$ and the quantum Langevin force $\sqrt{2\gamma}a^{in}(t)$, replacing the random force term $\sim F$ in classical Eq.\rf{1_1}, describe the interaction of the oscillator with the bath in Markovian approximation. The derivation of such terms from the system-bath interaction Hamiltonian can be found, for example, in \ct{LP} and in papers cited there;
\[
a^{in}(t)=\frac{1}{\sqrt{2\pi}}\int_{-\infty}^{\infty}a^{in}_{\omega}e^{-i\omega t}d\omega,
\]
where $a^{in}_{\omega}$ is Boze-operator of the bath mode: $[a^{in}_{\omega},a^{in+}_{\omega'}] = \delta(\omega + \omega')$. Coefficient $\sqrt{2\gamma}$ is chosen in the Langevin force term in order to provide Boze commutation relations for operators of the oscillator: $[a(t),a^+(t)] = 1$ -- as it is shown, for example, in \ct{CG}.

We suppose that the mean number of quanta in the bath $\left<a^{in+}(t)a^{in}(t)\right> >0$. The energy from the bath goes to the oscillator: the bath "pumps" the oscillator.

Classical nonlinear oscillator excited by regular force and described by  Eq.\rf{1_1} can have more than one stationary state near the resonance \ct{11}. Let us see wether many stationary states can appear in the quantum oscillator excited by random force and described by Eq.\rf{1_3a}. We'll find approximate stationary solution of Eq.\rf{1_3a} neglecting by fluctuations of energy $a^+a$ of the oscillator. In this approximation Eq.\rf{1_3a} reads:
\beq
    \dot{a} = [i(\delta + bn) - \gamma]a + \sqrt{2\gamma}a^{in}(t),  \lb{1_3}
\eeq
where $n = \left<a^+a\right>$ is dimensionless energy: average number of quanta in the oscillator. In the stationary case  $n$ is c-number, so that Eq.\rf{1_3} is linear equation respectively to fluctuating variable $a$, therefore Eq.\rf{1_3} can be solved by Fourier transform, as at standard analysis of fluctuations in linear systems \ct{10a}. In a difference with the linear analysis of fluctuations, $n$ in Eq.\rf{1_3} is unknown value, which itself depends on fluctuations and has to be determined.

Carrying out Fourier transforms in Eq.\rf{1_3} we come to relation between fourier-component operators:
\beq
    a_{\omega} = \frac{ \sqrt{2\gamma}a^{in}_{\omega}}{\gamma - i(\delta + \omega + bn)}, \hspace{0.5cm}
  o(t) = \frac{1}{\sqrt{2\pi}}\int_{-\infty}^{\infty}o_{\omega}e^{-i\omega t}d\omega, \hspace{0.5cm}
    o \equiv \{a, a^{in}\}. \lb{1_4}
\eeq
Using Eqs.\rf{1_4} we find stationary
\beq
    n=\left<a^+(t)a(t)\right> = \frac{1}{2\pi}\int_{-\infty}^{\infty}d\omega\int_{-\infty}^{\infty}d\omega'
    \left<a^+_{-\omega}a_{\omega'}\right>e^{(\omega + \omega')t}.    \lb{1_6}
\eeq
We insert expression \rf{1_4} for $a_{\omega}$ into Eq.\rf{1_6}. Pump bath Boze operators are delta-correlated: $\left<a^{in+}_{-\omega}a^{in}_{\omega'}\right> = n_{in}(\omega)\delta{(\omega + \omega')}$, where $n_{in}(\omega)$ is a number of quanta in pump bath mode of frequency $\omega$, so that:
\beq
       n  =  \frac{\gamma}{\pi}\int_{-\infty}^{\infty}\frac{n_{in}(\omega)d\omega}{\gamma^2 + (\delta + \omega + bn)^2}
    \equiv \frac{1}{2\pi}\int_{-\infty}^{\infty}n_{\omega}d\omega,  \lb{1_7}
\eeq
where $n_{\omega}$ is a noise power spectrum of the oscillator. Nonlinear integral equation \rf{1_7} determines $n$; in general Eq.\rf{1_7}  can be solved numerically. In the next Section we solve Eq.\rf{1_7} analytically for some particular $n_{in}(\omega)$. Note that assuming "white noise", when $n_{in}$ does not depend on $\omega$, we obtain that the resonance is absent: $n = n_{in}$ -- does not depend on $\omega$. Thus if we want to investigate the resonance, we have to consider the power spectrum of the random force of the finite width.

\section{Analysis of bistabilities}

We analyze Eq.\rf{1_7} for two examples of $n_{in}(\omega)$. Let us first approximate $n_{in}(\omega) \approx \left<n_{in}(\omega)\right> \equiv n_p$, that is an average number of quanta in one pump bath mode:
\beq
    n_{in}(\omega) = \left\{\begin{array}{cc} n_p\equiv \pi\gamma_p/\Gamma_p, & -\Gamma_p < \omega < \Gamma_p\\
    0 & \omega < -\Gamma_p,  \hspace{0.2cm}\omega > \Gamma_p \end{array}\right. . \lb{1_8}
\eeq
In Eq.\rf{1_9} we expressed  $n_p$ through the rate $\gamma_p$ of flux of quanta from the pump bath to the oscillator: $\gamma_p = (1/2\pi)\int_{-\infty}^{\infty}n_{in}(\omega)d\omega$.   We normalize Eq.\rf{1_7} and replace it by

\beq
       z  = \frac{1}{2\Gamma}\int_{-\Gamma}^{\Gamma}\frac{dx}{1 + (\Delta + x + Bz)^2},  \lb{1_9}
\eeq
where normalized energy $z$ and parameters are
\beq
    z = n\gamma/(2\gamma_p), \hspace{0.5cm}  \Delta = \delta/\gamma, \hspace{0.5cm}
    B = 2b\gamma_p/\gamma^2, \hspace{0.5cm} \Gamma = \Gamma_p/\gamma.  \lb{1_9a}
\eeq
Taking the integral in Eq.\rf{1_9} we obtain
\beq
      2\Gamma z  = \arctan{\left[\frac{2\Gamma}{1 +(\Delta  + Bz)^2 - \Gamma^2}\right]} + \eta \lb{1_10}
\eeq
with $\eta = 0$ for $\Gamma^2- (\Delta  + Bn)^2 < 1$, and $\eta = \pi$ for $\Gamma^2- (\Delta  + Bz)^2 > 1$.  By taking tangent from both sides of Eq.\rf{1_10} we came to
\beq
    \tan{(2\Gamma z)} = \frac{2\Gamma}{1 + (\Delta  + Bz)^2 - \Gamma^2}. \lb{1_11}
\eeq
Eq.\rf{1_11} determines  $z$ in unexplicit form, however using Eq.\rf{1_11} one can easily plot stationary $z(\Delta)$ and investigate necessary conditions for bistability. For that we express from Eq.\rf{1_11}  normalized detuning $\Delta$ as function of $z$:
\beq
    \Delta(z) = -Bz \pm \left[2\Gamma\cot{(2\Gamma z)} + \Gamma^2 - 1\right]^{1/2}.    \lb{1_12}
\eeq
For regular force, when $\Gamma \rightarrow 0$, Eq.\rf{1_12} came to
\beq
\Delta(z) = -Bz \pm \left(1/z - 1\right)^{1/2}.  \lb{1_12a}
\eeq
Using Eq.\rf{1_12} and taking $0<z<\pi/(2\Gamma)$ we plot $z(\Delta)$ in Fig.1.
%
%%%%%%%%%%%%%%%%%%%%%%%%%%%%%%%%%%%%%%%%%%%%%%%%%%%%%%%%%%%%%%%
%
\begin{figure}[h]
\bc \includegraphics[width=10cm]{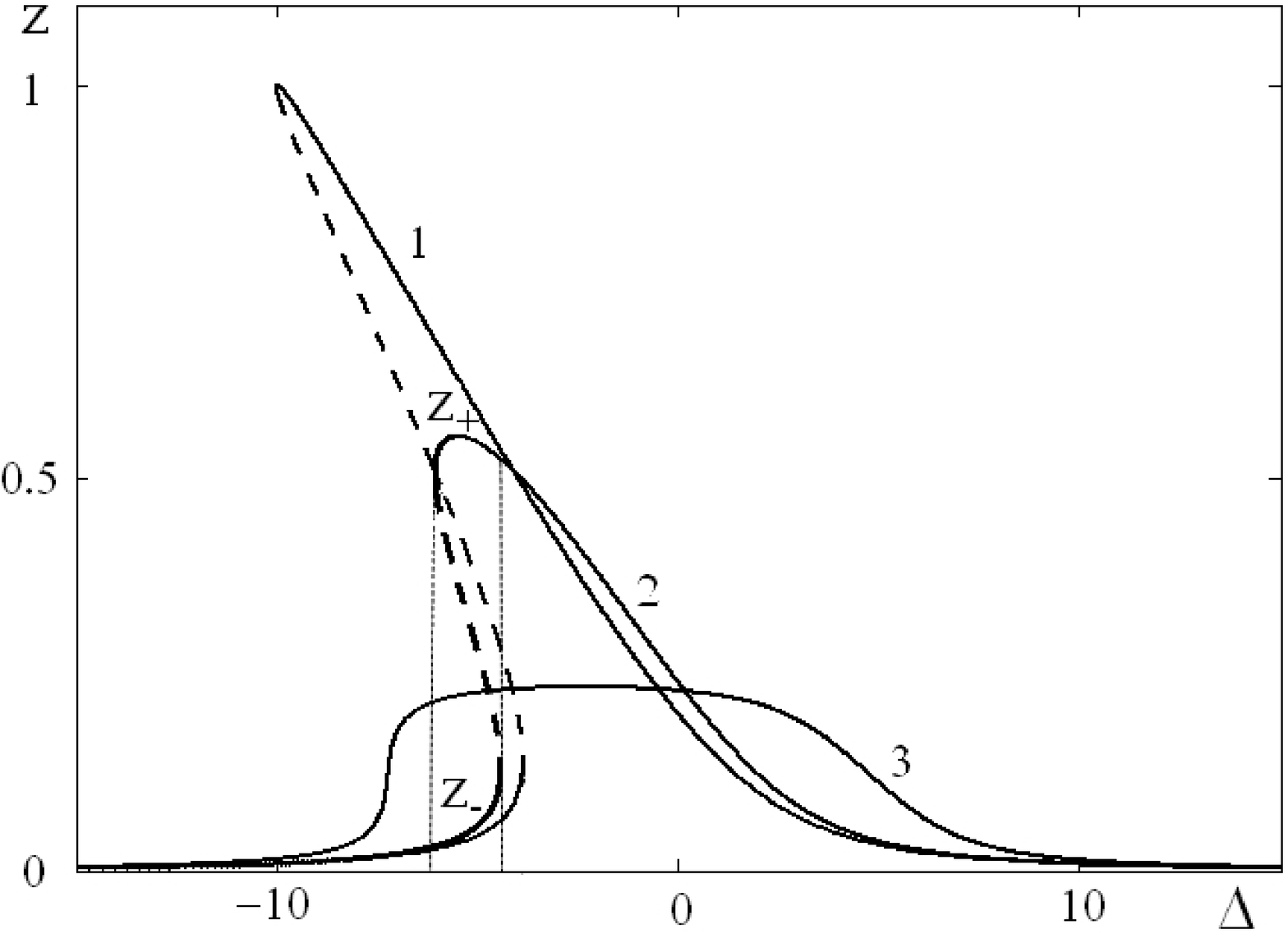}\\
\vspace{0.5cm}\baselineskip 0.5cm \parbox{16cm}{\small  {\bf Fig.1} {Normalized: energy $z$ of nonlinear oscillator versus detuning $\Delta$ for various values of half-width $\Gamma$ of the spectrum of random force exciting the oscillator. Normalized coefficient of nonlinearity $B = 10$. Curves 1, 2 and 3 are for $\Gamma = 0$ (regular force), $2$ and $6$, respectively. Dashed parts of $z(\Delta)$ curves correspond to unstable stationary solutions. The region of $\Delta$ between vertical dashed lines corresponds to bistability in curve 2 where $z_-$ is the lower branch, $z_+$ is the upper branch.}} \ec
\end{figure}
%
%%%%%%%%%%%%%%%%%%%%%%%%%%%%%%%%%%%%%%%%%%%%%%%%%%%%%%%%%%%%%%%
%
One can see from Fig.1 that fluctuations of exciting force broad the oscillator spectrum. However for given value of $B=10$ three stationary solutions exist for the same $\Delta$, if dimensionless  width $\Gamma$ of the random force power  spectrum is not too large: one can observe  bistability in the curve 2 for $\Gamma = 2$. For larger $\Gamma$, as $\Gamma = 6$ for the curve 3, bistability disappears.

We can find necessary conditions for bistability in nonlinear resonance by using Eq.\rf{1_12} and applying approach of catastrophe theory \ct{12}. It is clear from Fig.1 that bifurcation points, when the number of stationary $z$ is changed from 1 to 3, correspond to $d\Delta/dz = 0$, which is the same as
\beq
    B = \frac{2\Gamma^2[1+ \cot^2{(2\Gamma z)}]}{\left[2\Gamma\cot{(2\Gamma z)} +
%d
    \Gamma^2-1\right]^{1/2}}. \lb{1_13}
\eeq
Inserting $z$ from domain $0<z<\pi/(2\Gamma)$ into Eqs.\rf{1_12} and \rf{1_13} we find regions of bistability in parameter space $\{\Delta, B, \Gamma\}$, see Fig.2.
%
%%%%%%%%%%%%%%%%%%%%%%%%%%%%%%%%%%%%%%%%%%%%%%%%%%%%%%%%%%%%%%%
%
\begin{figure}[h]
\bc \includegraphics[width=10cm]{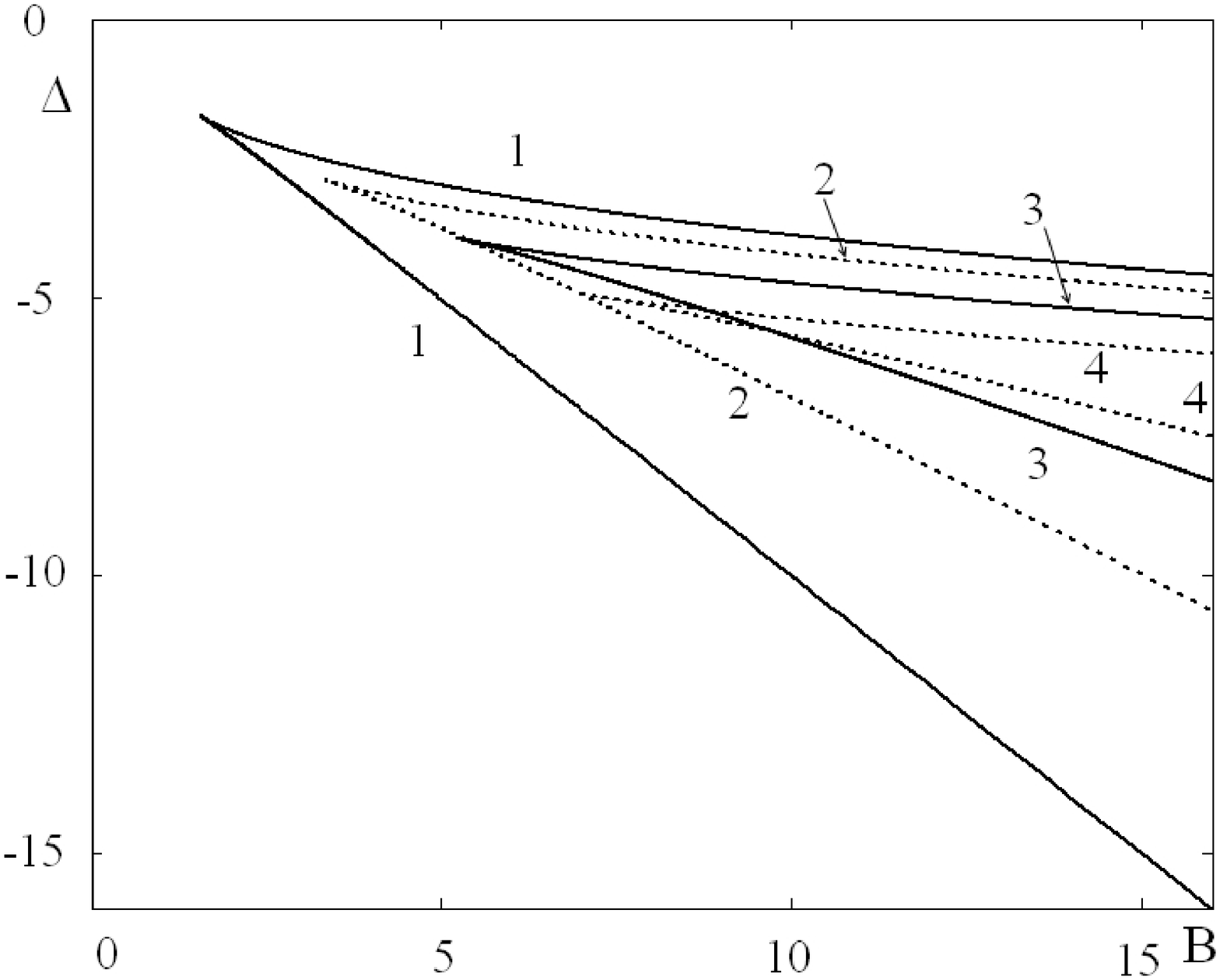}\\
\vspace{0.5cm}\baselineskip 0.5cm \parbox{16cm}{\small  {\bf Fig.2} {Regions of bistability in $B, \Delta$ parameter space for various values of $\Gamma$. Bistability exists for $\Delta$, $B$ from the region between curves 1 for $\Gamma =0$ (regular force); between curves 2, 3 and 4 for $\Gamma = 1.5$, $2.5$ and $3.5$, respectively. Maxima $B$ in each curve (where $d\Delta/dB$ does not exist) correspond to $B = B_{bif}$.}} \ec
\end{figure}
%
%%%%%%%%%%%%%%%%%%%%%%%%%%%%%%%%%%%%%%%%%%%%%%%%%%%%%%%%%%%%%%%
%
As larger is $\Gamma$, i.e. as noisy is the exciting force, as larger is the value of $B$ and  $b\gamma_p \sim B$ necessary for bistability. With the increase of $\Gamma$ bistability appears at lager  $|\Delta|$ and in narrower interval of $\Delta$.

According to \ct{12}, bistability appears if $B>B_{bif}$, where $B_{bif}$ is determined from condition $d^2\Delta/dz^2 = 0$ at $d\Delta/dz = 0$, i.e. when Eq.\rf{1_13} is true. Demanding $d^2\Delta/dz^2 = 0$ which is the same as $dB/dz =0$, where $B(z)$ is given by Eq.\rf{1_13}, we obtain
\beq
    \cot(2\Gamma z) =
    \frac{1}{3\Gamma}\left\{1 - \Gamma^2 + \left[(1-\Gamma^2)^2 + 3\Gamma^2\right]^{1/2}\right\}.    \lb{1_14}
\eeq
Inserting $\cot(2\Gamma z)$ from Eq.\rf{1_14} into Eq.\rf{1_13} we find "bifurcation" curve $B_{bif}(\Gamma)$, which separates the region of bistability from the region with single stationary solution in $B, \Gamma$ parameter space, see Fig.3.
%
%%%%%%%%%%%%%%%%%%%%%%%%%%%%%%%%%%%%%%%%%%%%%%%%%%%%%%%%%%%%%%%
%
\begin{figure}[h]
\bc \includegraphics[width=8cm]{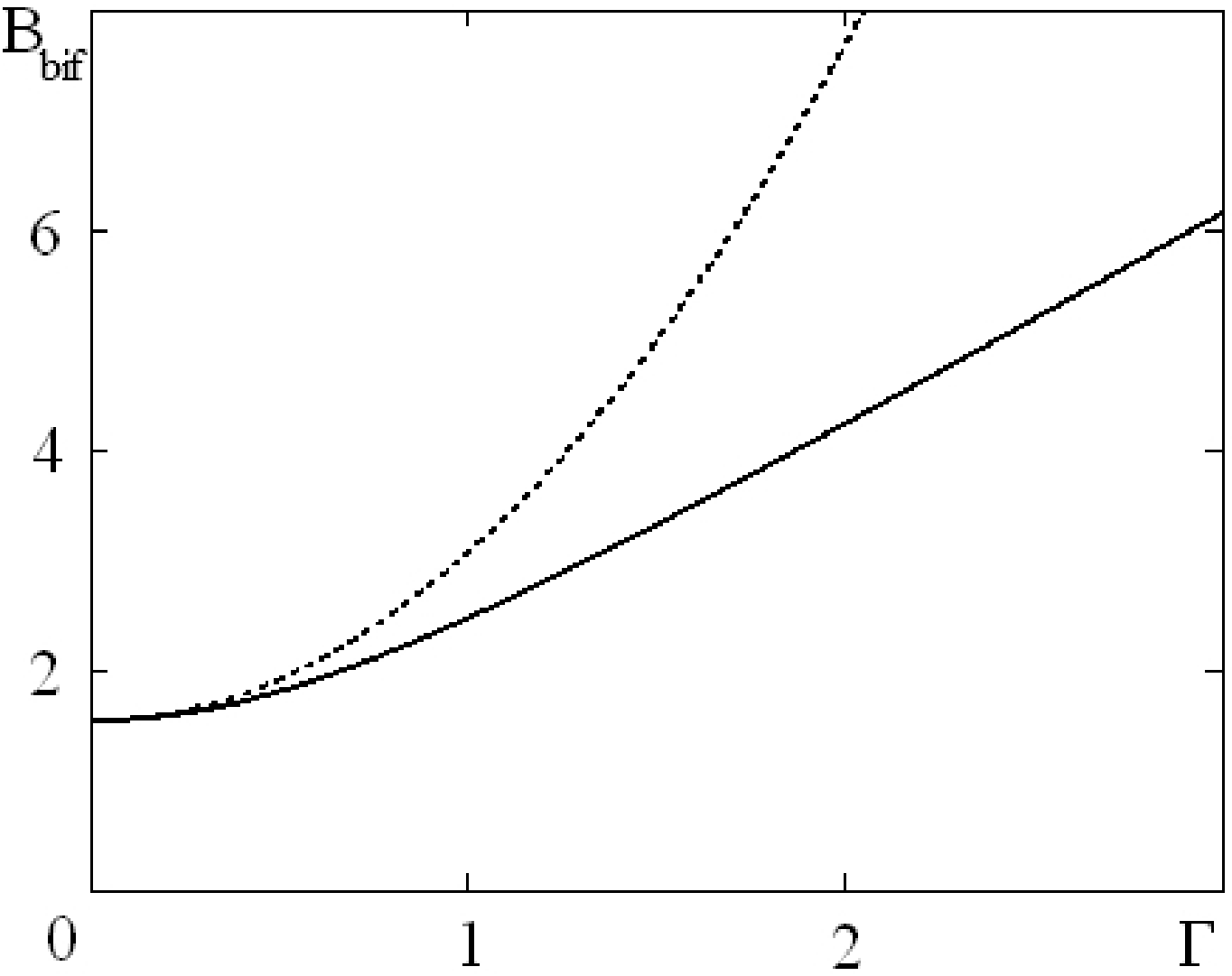}\\
\vspace{0.5cm}\baselineskip 0.5cm \parbox{16cm}{\small  {\bf Fig.3} {Bistability exists for values of $B$ above the solid curve for restangular noise power spectrum of random force and above the dashed curve for Lorenz noise power spectrum of the force.}} \ec
\end{figure}
%
%%%%%%%%%%%%%%%%%%%%%%%%%%%%%%%%%%%%%%%%%%%%%%%%%%%%%%%%%%%%%%%
%
One can see from Fig.3 that $B_{bif}(\Gamma)$ is almost a straight line for $\Gamma >1$. Thus the minimum value of $b\gamma_p \sim B$ necessary for bistability is  $b\gamma_p \sim \Gamma$.

Stationary solutions shown by dashed parts of $z(\Delta)$ curves in Fig.2 corresponding to $dz/d\Delta >0$ are unstable. Indeed, using Eq.\rf{1_3} we write equation of motion for $n$:
\beq
\frac{dn}{dt} \equiv \left< \frac{da^+}{dt}a + a^+\frac{da}{dt}\right> = -2\gamma n + \sqrt{2\gamma}(\left<a^{in+}a\right> + \left<a^+a^{in}\right>). \lb{1_13a}
\eeq
Inserting in Eq.\rf{1_13a} $a(t)$ and $a^{in}(t)$ expressed through fourier components as in Eq.\rf{1_4}, replacing $n$ by $z$ and coming  to normalized  parameters shown by Eq.\rf{1_9a} we obtain
\beq
\frac{dz}{dt} = -2\gamma \left[z - \int_{-\Gamma}^{\Gamma}\frac{dx}{1 + (\Delta + x + Bz)^2}\right]. \lb{1_13b}
\eeq
Linear stability analysis of Eq.\rf{1_13b} shows that $z(t)$ is unstable if $d z/d\Delta >0$, where $z$ is the stationary solution of Eq.\rf{1_13b} inexplicitly determined by Eq.\rf{1_12}.

Analysis of stability of the upper and the lower branches of $z(\Delta)$ curves, as $z_{\pm}$ in Fig.1, is not so straightforward. When the system is in the bistability region, the fluctuating force can, with some probability, produce large fluctuation, which switches the system from one stationary state to another \ct{15}, \ct{16}. Thus stationary states in the bistabiliy region are meta-stable. The system stays only finite time $\tau_l$ in such "quazi-stationary" states; $\tau_l$ depends on how far are "quazi-stationary" states from each other and on fluctuating force spectrum and strengths. Suppose that $z_+>z_-$ are normalized energies of quaz-stationary states: they correspond to the upper ($z_+$) and the lower ($z_-$) brunch of, for example, curve 2 in Fig.1 in the region of bistability. If the normalized fluctuation of the oscillator energy $\delta z \equiv (\left< z^2 \right> - \left< z \right>^2)^{1/2} > z_+ - z_-$ bistbility will be, obviously, destroyed. In such case $\tau_l < \gamma^{-1}$ -- typical time of the relaxation of the oscillator to its stationary state. By considering Eq.\rf{1_13a} as Heisenberg equation for operator $\hat{n} = a^+a$ one can calculate fluctuations of $\hat{n}$ and find $\delta z$. Fluctuations of the oscillator energy will lead to narrowing of the bistability regions respectively to ones shown in Fig.2 and to metastability (final lifetime) of oscillator stationary states. We will carry more detailed analysis of metastable states of our quantum system and their properties in the future. 

Suppose now that the noise spectrum of pump bath is very broad: $\Gamma \gg 1$. In this case we can approximate $(\gamma/\pi)[\gamma^2 + (\delta + \omega + bn)^2]$ in Eq.\rf{1_7} by Dirac delta-function, so that
\beq
   n \approx \frac{\gamma}{\pi}\lim_{\gamma\rightarrow 0}\int_{-\infty}^{\infty}\frac{n_{in}(\omega)d\omega}{\gamma^2 + (\delta + \omega + bn)^2} = n_{in}(-\delta - bn).\lb{1_14a}
\eeq
For example, if we take Lorenz power spectrum $n_{in}(\omega) = (2\gamma_p\gamma\Gamma)[\omega^2 + (\gamma\Gamma)^2]^{-1}$ then Eq.\rf{1_14a} in normalized quantities given by Eqs.\rf{1_9a} became $z = \Gamma[(\Delta + Bz)^2 + \Gamma^2]^{-1}$. Replacing $z' = \Gamma z$, $B' = B/\Gamma^2$ and $\Delta' = \Delta/\Gamma$ we obtain
\beq
   \Delta'(z') = -B'z' \pm \left[1/z' - 1\right]^{1/2},    \lb{1_15}
\eeq
that is the same (apart of notations) as Eq.\rf{1_12a} for nonlinear oscillator excited by regular force. One can find bistability conditions $B' >8\sqrt{3}/9$, therefore $B > B_{bif} = (8\sqrt{3}/9)\Gamma^2$, that is shown in Fig.3 by dashed line. In Fig.3 we shifted the curve $B_{bif} = (8\sqrt{3}/9)\Gamma^2$ up in order to be consisted with $B = B_{bif} = 8\sqrt{3}/9$ at $\Gamma = 0$. One can see that conditions for bistability for Lorenz noise power spectrum are qualitatively different than conditions for restangular specrtum: in the first case $B > B_{bif} \sim \Gamma^2$ in the last case $B > B_{bif} \sim \Gamma$.

With the help of Eq.\rf{1_14} one can investigate conditions for bistability at $\Gamma \gg 1$ for another power spectrum of excitation forces, for example for gaussian spectrum etc.

\section{Conclusion}
Using an example of nonlinear oscillator excited by random force we demonstrated approximate method of analysis of quantum nonlinear systems with strong noise. We neglected by fluctuations in amplitude of the oscillator preserving fluctuations in its phase. This  is natural first-order approximation for studying an oscillating system excited by noisy bath, when the mean energy of oscillations is not zero, while the phase fluctuates are on $[0,2\pi]$ interval. Well-known example of such  quantum system is a laser described by our method in \ct{9,10}. In our method the oscillator power spectrum $n_{\omega}$ depends on the oscillator mean energy $n$ so that $n \equiv (2\pi)^{-1}\int_{-\infty}^{\infty}n_{\omega}(n)d\omega$ is nonlinear integral equation for $n$. We solved this equation for some particular cases. Thus, in a difference with usual linear methods of analysis of the noise \ct{10a}, we take into account the influence of the noise to the stationary state of the system.

Here we found necessary conditions for bistability: regions of parameters, where more than one stationary solution exist for quantum nonlinear oscillator driven by random force with non-white spectrum of fluctuations. Fluctuations of exciting force broad the resonance, more than one stationary solution exists at the resonance sideband, if dimensionless nonlinearity parameter $B$ is large; $B \sim \gamma_pb$, where $\gamma_p$ is the excitation rate and $b$ is coefficient of nonlinearity in sec$^{-1}$. Necessary condition for bistability is $B>B_{bif}\sim \Gamma$, where $\Gamma$ is a width of restangular power spectrum of the random force, or $B>B_{bif}\sim \Gamma^2$ -- for the case of Lorenz power spectrum of the force. Thus necessary bistability conditions are substantially different for different noise power spectrums.

Here we did not take into account amplitude fluctuations. When amplitude fluctuations are of the order of the mean value of the energy of the oscillator in the region of bistability, the bistability will be destroyed: the oscillator can't stay in the lower or in the upper stationary states of the bistability curve. We do not study here dynamics of switchings between states.  In the future we'll estimate the contribution of amplitude fluctuations, which, at first approximation, can be done with the help of Heisenberg equation, as Eq.\rf{1_13a}, for the operator of energy. In order to find sufficient conditions for the bistability in this quantum system more detailed analysis has to be done, as it was, for example, in \ct{15} and \ct{16} for the classical case.

Similar way, with our method one can study various interesting phenomena in dynamic of quantum nonlinear systems with noise. For example, one can consider a combination of broad and narrow banded random force excitation of nonlinear oscillator, as it is for the case stochastic resonance \ct{1,2,3} also in the three-level atomic optical bistability (AOB) systems \ct{06} -- \ct{010}. Nonlinear oscillator considered here is similar with the molecular vibration mode excited by short and, therefore, spectrally broad laser pulse. So that our results can be used for estimations of conditions of bistability at laser excitation of molecules in selective laser chemistry \ct{13}.

\end{document}